\documentclass[12pt]{article}
\usepackage{a4wide,epsfig,psfrag,amsmath,amssymb,cite,scalefnt}

\newcommand{\gsim}{\;\rlap{\lower 3.5 pt \hbox{$\mathchar \sim$}} \raise 1pt
 \hbox {$>$}\;}
\newcommand{\lsim}{\;\rlap{\lower 3.5 pt \hbox{$\mathchar \sim$}} \raise 1pt
 \hbox {$<$}\;}

\begin{document}


\title{\vskip-3cm{\baselineskip14pt
    \begin{flushleft}
      \normalsize SFB/CPP-08-74\\
      \normalsize TTP08-43
  \end{flushleft}}
  \vskip1.5cm
  ${\cal O}(\alpha\alpha_s)$ corrections to the $\gamma t\bar{t}$ vertex at
  the top quark threshold
}
\author{\small Yuichiro Kiyo, Dirk Seidel and Matthias Steinhauser\\[1em]
  {\small\it Institut f{\"u}r Theoretische Teilchenphysik,
    Universit{\"a}t Karlsruhe (TH)}\\
  {\small\it Karlsruhe Institute of Technology (KIT)}\\
  {\small\it 76128 Karlsruhe, Germany}}

\date{}

\maketitle

\thispagestyle{empty}

\begin{abstract}
We compute the last missing piece of the 
two-loop ${\cal O}(\alpha \alpha_s)$ corrections
to $\gamma t \bar{t}$ vertex at the $t \bar{t}$ threshold 
due to the exchange of a $W$ boson and a gluon. This contribution
constitutes a building 
block of the top quark threshold production cross section at electron 
positron colliders.
\medskip

\noindent
PACS numbers:12.38.Bx,12.15.-y,12.38.-t,14.65.Ha

\end{abstract}

\thispagestyle{empty}


\newpage


\section{Introduction}

The top quark pair production close to the threshold 
is an important process at a future International Linear
Collider (ILC). It can be used to 
determine top quark properties, like the mass $m_t$ and the width 
$\Gamma_t$, but also the strong coupling $\alpha_s$ with high
precision. This is in particular true for $m_t$ where an uncertainty
below 100~MeV can be obtained from a threshold scan of 
the cross section \cite{Martinez:2002st}.
 
The feasibility of such high-precision measurements
requires a theory prediction of the total cross section
$\sigma(e^+ e^-\rightarrow t\bar{t})$ 
with high accuracy (preferably $\delta\sigma/\sigma\leq 3\%$).
Current estimates based on (partial)
next-to-next-to-leading logarithmic
(NNLL) order~\cite{Hoang:2003xg,Pineda:2006ri} and (partial)
next-to-next-to-next-to-leading order
(NNNLO)~\cite{Beneke:2008ec,Beneke:2007pj} 
QCD corrections lead to an uncertainty of the order of 10\%.

In order to reach a theory goal of $\delta\sigma/\sigma\leq 3\%$ 
it is necessary to 
include in the prediction next to the one-loop electroweak
corrections, which are known since quite some time~\cite{Guth:1991ab}
(see also~\cite{Hoang:2006pd}), also higher order effects.
The evaluation of ${\cal O}(\alpha \alpha_s)$ corrections
has been started in Ref.~\cite{Eiras:2006xm}, where the two-loop mixed 
electroweak and QCD corrections to
the matching coefficient of the vector current has been computed due to
a Higgs or $Z$ boson exchange in addition to a gluon.
The current paper continues this enterprise 
and provides a result of ${\cal O}(\alpha\alpha_s)$ for the two-loop 
vertex diagrams mediated by a $W$ boson and gluon 
exchange.\footnote{Of
  course, in addition to the gauge boson also the corresponding  
  Goldstone boson is taken into account.}
This completes the vertex corrections of order $\alpha\alpha_s$ --- a
building block for the top quark production cross section.
Assuming the (numerically well justified) power counting
$\alpha \sim \alpha_s^2$ one can see that these corrections are 
formally of NNNLO.

In order to complete the matching corrections of order
$\alpha\alpha_s$ also the two-loop box diagrams 
contributing to $e^+ e^-\to t\bar{t}$ have to be considered.
Actually, only the proper combination of the box, vertex and
self-energy contributions (the latter can, e.g., be found in 
Refs.~\cite{Kniehl:1989yc,Djouadi:1993ss})
forms a gauge independent set. 

The remainder of the paper is organized as follows: 
In the next Section we introduce our notation and derive the cross
section formula for $e^+e^-\rightarrow t\bar{t}X$ near the $t\bar{t}$ 
threshold. We present a general formula which includes all radiative
corrections of the Standard Model (SM). 
In Section~\ref{sec::tech} we discuss some
technical details of the two-loop computation and 
in Section~\ref{sec::gammatt} we concentrate on the ${\cal  O}(\alpha\alpha_s)$ 
corrections to the $\gamma t\bar{t}$ vertex and present 
our results. Section~\ref{sec::concl} contains
our conclusions. Additional useful material concerning the one-loop
expressions can be found in the Appendix.


\section{\label{sec::xs}Threshold cross section}

The production cross section for the process $e^+e^-\to t\bar{t}X$
near threshold consists of helicity amplitudes and the hadronic part. 
The former correspond to the high-energy production amplitude of a
top quark pair, the latter to the QCD
bound-state dynamics of the produced $t\bar{t}$ pairs exchanging gluons 
to form a resonance.
The cross section can be cast in the form (for left-landed positrons
and right-handed electrons),
\begin{eqnarray}
  R\left(e_{L}^+ e_{R}^-  \rightarrow t\bar{t} X\right)
  &=&
   \frac{8\pi}{s}\,
   {\rm Im}
   \big[\, \left(h_{R,\,V}\right)^{\,2} H_{V}
           +\left(h_{R,\,A}\right)^{\,2} H_{A} \big]\,,
  \label{eq:XSection}
\end{eqnarray}
where $s$ is the square of the center-of-mass energy and
$R(e_{L}^+ e_{R}^-  \rightarrow t\bar{t} X)$ is the
cross section normalized to $\sigma(e^+e^-\rightarrow \mu^+\mu^-)=(4\pi\alpha^2)/(3s)$. 
The first subscript of $h$ refers to helicity 
of the electron, and the second one to the
vector ($J_{V}^\mu=\bar{\psi}\gamma^\mu\psi$) or 
axial-vector coupling ($J_{A}^\mu=\bar{\psi}\gamma^\mu\gamma_5\psi$)
of the gauge bosons to the top quark current.\footnote{The notation
  is basically adapted from Ref.~\cite{Guth:1991ab}, however, 
  we added a second subscript $F=V,A$ to incorporate the axial-vector
  coupling of the $Zt\bar{t}$ vertex (see Eq.~(\ref{eq:hamp})).}
For $e^+_R e^-_L$ in the initial state a similar expression 
is obtained by replacing R by L in Eq.~(\ref{eq:XSection}).
In the SM the helicity amplitudes for $e^+_R e^-_R$ and $e^+_Le^-_L$ 
are proportional to $m_e/M_W\sim 10^{-5}$ and are thus negligible.

In the cross section formula (\ref{eq:XSection}) ``${\rm Im}$'' refers
to those cuts which correspond to the $t\bar{t}X$ final
state.\footnote{From the theoretical point of view one has a pure 
  $t\bar{t}$ final state up to NNLO in QCD. Starting from NNNLO one has
  to include the real emission of a gluon.
  Once the electroweak sector is considered, final states 
  like $(bW^+)\bar{t}$ needs to be included, where $(bW^+)$ has 
  an invariant mass $p_{bW}^{\,2}$ in a range 
  $|p_{bW}^{\,2}-m_t^{\,2}| \lesssim m_t \Gamma_t$.}
This means that we have to select special cuts which correspond to 
the final state we are interested in. This requires a 
dedicated study incorporating the experimental setup.
For the one-loop electroweak correction 
this treatment was performed in \cite{Hoang:2004tg}.
In this paper we will not pursue this problem further (see also the
discussion in the Conclusions). 

The hadronic part is described by non-relativistic QCD
(NRQCD)~\cite{Bodwin:1994jh}. For our purpose it is sufficient
to re-write the vector current $J_{V}^\mu$ and the 
axial-vector current $J_{A}^\mu$
in terms of two-component NRQCD spinor 
fields $\psi$, $\chi$, which correspond to non-relativistic top and 
anti-top quarks, respectively. This yields the following NRQCD currents 
\begin{eqnarray}
  && j_V^{\,i} = \psi^\dag\,\sigma^i\chi, ~~~
  j_{V}^{(1/m^2),\, i}=-\frac{1}{6 m_t^2}\psi^\dag\,\sigma^i (i{\vec D}\,)^2\,
  \chi, ~~~ 
  j_A^{\,i} = \frac{1}{2m_t}\, \psi^\dag\,
  [\sigma^i,\left({\vec\sigma}\,i {\vec D}\,\right)]\,\chi\,.
  \label{eq:jvandja}
\end{eqnarray}
With the help of the NRQCD equation of motion for top and anti-top
quarks ($\psi^\dag \sigma^i\,{\vec D\,}^2\chi=m_t\,i \partial_0
(\psi^\dag\sigma^i \chi)$) the $1/m^2$-suppressed vector current can be
re-expressed in terms of $j_V$. Thus our matching relation between SM
and NRQCD currents are given by 
\begin{eqnarray}
  &&
  J_{V}=e^{2im_tx_0}\left(c_v-\frac{d_v}{6m_t}i\partial_0\right) \,j_{V}, ~~~
  J_{A}=e^{2im_tx_0}c_a\,j_{A},
  \label{eq:Jmatch}
\end{eqnarray}
with $c_v=d_v=c_a=1$ at tree level. 
The hadronic part is defined by the current correlation function
\begin{eqnarray}
  H_{F}
  &=& i\, \sum_{k}\int{\rm d}x \, e^{iE x^0}\,
  \langle \Omega\,|\, {\rm T} \,j_{F}^{\,k\, \dag} (x)\, j_{F}^{\, k}(0)\,
  |\Omega\rangle ~~~~(F=V, A), 
  \label{eq:H}
\end{eqnarray}
where $E=\sqrt{s}-2m_t$ and $|\Omega\rangle$ is the NRQCD vacuum state.

The evaluation of $H_F$ requires to integrate out the low-energy modes 
of QCD, the soft, potential and ultrasoft gluons contained in NRQCD 
\cite{Luke:1997ys,Beneke:1997zp}.
For the top quark system this can be done perturbatively. In a first
step one integrates out the soft and potential gluons which results 
in the effective field theory Potential
NRQCD~\cite{Pineda:1997bj,Brambilla:1999xf}. 
The corresponding Lagrangian is known to
NNNLO~\cite{Kniehl:2002br}.\footnote{The only missing constant in
  Ref.~\cite{Kniehl:2002br} is related to the three-loop static
  potential where recently the fermion corrections became
  available~\cite{Smirnov:2008pn}.}
To integrate out nonrelativistic top and anti-top quark fields 
the Rayleigh-Schr\"odinger perturbation theory can be applied as was 
initiated in Ref.~\cite{Fadin:1987wz} and performed to NNNLO for
$H_V$ and $H_A$ in Refs.~\cite{Beneke:2008ec} and~\cite{Penin:1998mx},
respectively. Integrating out the ultrasoft gluon 
was completed recently in Ref.~\cite{Beneke:2007pj}.
For the details of these steps we refer the reader to the original
papers and references cited therein (see also
Refs.~\cite{Kniehl:1999ud,Manohar:2000kr,Kniehl:2002yv,Hoang:2003ns,Penin:2005eu}).

In this paper we restrict ourselves to hard loop 
corrections to the production cross section, namely the corrections 
being parameterized as $h_{I,F}$. The tree-level
expression\footnote{We include the effect due to $j_V^{(1/m^2)}$ with
  $d^{\rm tree}_v=1$  into  $h_{I,V}^{(1/m^2)}$ for convenience, see
  Eq.~(\ref{eq:EWcorr2h}).}
of helicity amplitude $h_{I,F}$ is given by 
\begin{eqnarray}
  h_{I,V}^{\rm tree} &=& 
  Q_e Q_t 
  + \frac{s \,\beta_I^{\,e} \,\beta_V^{\,t}}{s-M_Z^2}~~~
  {\rm with}~~ \beta_V^{\,t}=\frac{\beta_R^{\,t}+\beta_L^{\,t}}{2},
  \nonumber \\
  \beta_{I}^f &=&\frac{(T_3)_{\,f_I}-s_w^2 Q_f}{s_w c_w}  ~~~~(I=L/R),
\label{eq:hamp}
\end{eqnarray}
where the $\beta_{I}^{\,f}$ is the coupling of a fermion ($f=e, t$) 
to the $Z$ boson, $s_w$  is the sin of the weak-mixing
($c_w^2=1-s_w^2$), 
and electric and iso-spin charges for top quark and electron are
given by 
\begin{eqnarray}
  Q_e=-1, ~~ Q_t=2/3, ~~ 
  (T_3)_{\,t_L}=1/2, ~~(T_3)_{\,e_L}=-1/2,~~
  (T_3)_{\,f_R}\equiv 0.
  \label{eq:charges}
\end{eqnarray}
In the following the abbreviation $T_3^f\equiv (T_3)_{f_L}$ will be used. 
$h_{I,A}$ can be obtained by substituting $\beta_V^t$ by 
$\beta_A^{\,t}=(\beta_R^t-\beta_L^t)/2$ in formula~(\ref{eq:hamp}).

Let us now explain how hard loop corrections within SM can be 
incorporated into the helicity amplitude $h_{I, F}$. To this end
we organize the corrections as
\begin{eqnarray}
  h_{I,V} &=& \left(h_{I,V}^{\rm tree}+h_{I,V}^{(1/m^2)}\right)
  + h_{I,V}^{(1,0)} + h_{I,V}^{(0,1)} + h_{I,V}^{(1,1)}\,,
  \nonumber \\
  h_{I,A} &=& h_{I,A}^{\rm tree}+h_{I,A}^{(0,1)}\,,
  \label{eq:EWcorr2h}
\end{eqnarray}
where the $h_{I,F}^{\rm tree}$ and $h_{I,V}^{(1/m^2)}$ (due to the
$j_V^{(1/m^2)}$) are the tree-level contributions, and $h_{I,F}^{(i,j)}$
incorporate the contributions from radiative corrections (the
superscript $(i,j)$ denotes the electroweak- and QCD-loop order, respectively).
As one can see from the expression of the axial-vector current
$j_{A}$, $H_A$ is suppressed by ${\vec D\,}^2/m_t^2\sim E/m_t$. 
Thus one-loop QCD corrections to $h_{I,A}$ correspond to NNNLO
effects. Hard QCD corrections to the 
$\gamma t\bar{t}$ and $Zt\bar{t}$ vertices modify the matching
coefficients $c_{v,a}$ at loop level. We absorb these effects into 
helicity amplitudes and obtain (using
$h_{I, V}^{(1/m^2)}=-h_{I,V}^{\rm tree} \,\,E/(6m_t)$)
\begin{eqnarray}
  h_{I, V}^{(0,1)}= \big(c_v^{\,(1)}-\frac{E\,d_v^{\,(1)}}{6m_t}\big)\, h_{I,V}^{\rm tree}\,, ~~~
  h_{I, A}^{(0,1)}= c_a^{\,(1)}\, h_{I,A}^{\rm tree}\,,
\label{eq:cvcorr2h}
\end{eqnarray}
with $c_v^{\rm (i)},\, d_v^{\rm (i)}$ being the $i$-loop contribution to the 
matching coefficients.
For the purpose of this paper only the one-loop
contribution $c_v^{\, (1)}$ is needed (see below for explicit expressions).
Let us mention that the two-loop QCD corrections have been evaluated
in Refs.~\cite{Czarnecki:1997vz,Beneke:1997jm} and the three-loop
corrections induced by a light quark loop in Ref.~\cite{Marquard:2006qi}.


\section{\label{sec::tech}Technical details of the two-loop calculation}

\begin{figure}[t]
  \begin{center}
    \includegraphics[width=.3\textwidth]{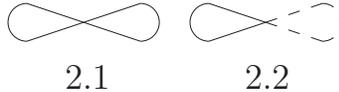}
    \caption{Two-line MIs. The solid and dashed lines correspond to
      propagators with mass $m_t$ and $M_W$, respectively.}
    \label{fig::MI2}
  \end{center}
\end{figure}

\begin{figure}[t]
  \begin{center}
    \includegraphics[width=.8\textwidth]{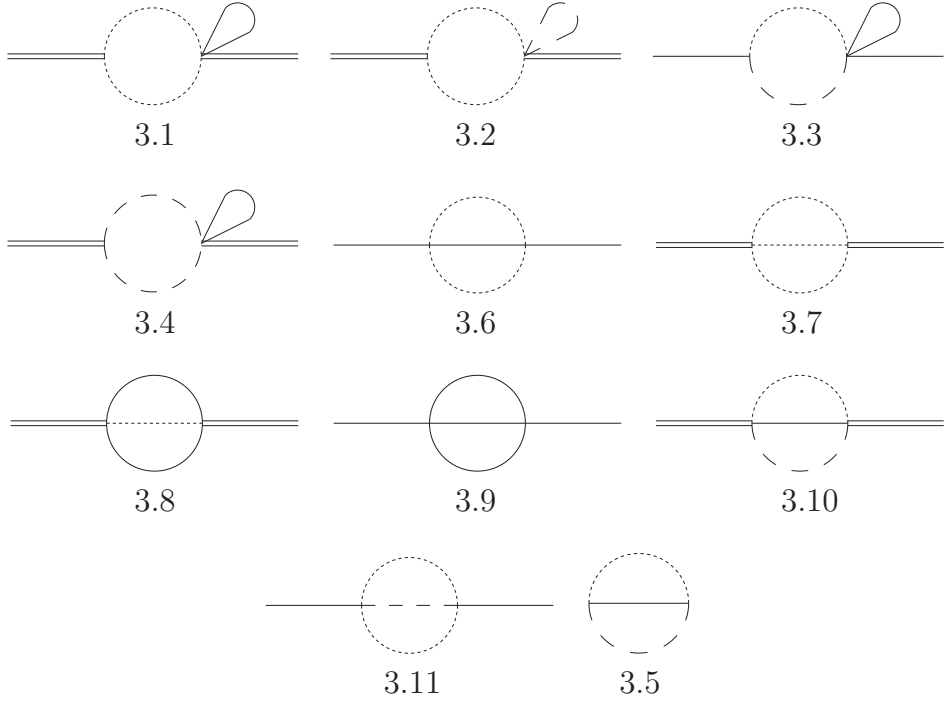}
    \caption{Three-line MIs. The solid, dashed and dotted lines correspond to
      propagators with mass $m_t$, $M_W$ and 0, respectively. Single external lines are on the mass
      shell with mass $m_t$ whereas double external line have mass $2m_t$. The topology denoted by
      3.10 contains three master integrals, topology 3.11 contains two MIs.}
    \label{fig::MI3}
  \end{center}
\end{figure}

\begin{figure}[t]
  \begin{center}
    \includegraphics[width=.8\textwidth]{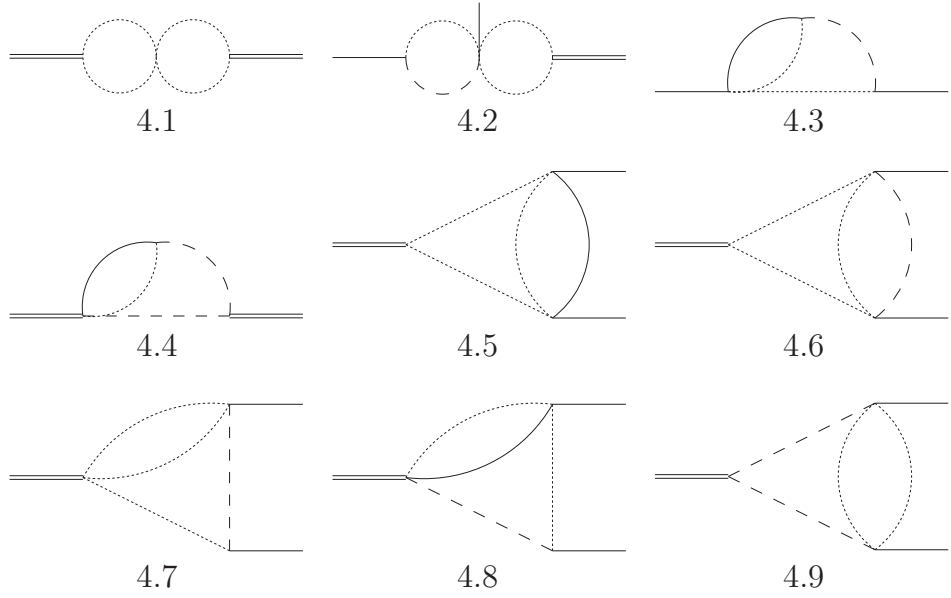}
    \caption{Four-line MIs. For the notation we refer to Fig.~\ref{fig::MI3}.}
    \label{fig::MI4}
  \end{center}
\end{figure}

\begin{figure}[t]
  \begin{center}
    \includegraphics[width=.6\textwidth]{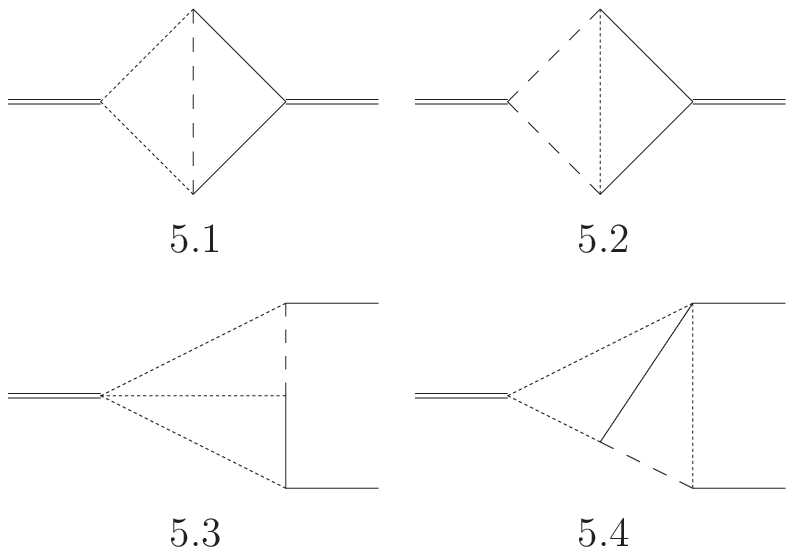}
    \caption{Five-line MIs. For the notation we refer to Fig.~\ref{fig::MI3}.}
    \label{fig::MI5}
  \end{center}
\end{figure}

Let us in this Section provide some technical details about the
evaluation of the two-loop diagrams. They are generated with {\tt
  QGRAF}~\cite{Nogueira:1991ex} and further processed with {\tt q2e} and {\tt
  exp}~\cite{Harlander:1997zb,Seidensticker:1999bb}. The reduction of the
integrals is performed with the 
program {\tt crusher}~\cite{PMDS} which implements the Laporta
algorithm~\cite{Laporta:1996mq,Laporta:2001dd}. 
We arrive at 29 master integrals (MI) which are depicted in
Figs.~\ref{fig::MI2}--\ref{fig::MI5}. All diagrams occur with 
the propagators raised to power one. Note that there are two more MIs
of type 3.10: one with a squared top quark propagator and one with a
squared massless propagator. Similarly, an additional MI
arises from type 3.11 with a squared massless propagator.

We refrain from presenting the explicit results for all MIs in this 
paper but provide them in form of a {\tt Mathematica} file\footnote{The file
  is available from {\tt http://www-ttp.particle.uni-karlsruhe.de/Progdata/ttp08/ttp08-43.}
}
{\tt MIttewW.m}
using the conventions as defined in Eq.~(\ref{eq::A0}) which
corresponds to the one-loop tadpole integral.
In all results presented in this file a factor
$\left(\mu^2/m_t^2\right)^{2\epsilon}$ 
with $d=4-2\epsilon$ has to be multiplied. 

Some MIs factorize into one-loop integrals or contain only one dimensionful
scale. Most of these integrals are available in the literature and can, e.g., be
found in
Refs.~\cite{Scharf:1993ds,Fleischer:1998dw,Seidel:2004jh,Kalmykov:2006pu,Eiras:2006xm,Piclum:2007an}.

As we will see in Section~\ref{sec::gammatt} a rapid convergence is observed
if one considers an expansion of the matching coefficient in the quantity
$z=M_W^2/m_t^2$. 
For this reason we evaluate the two-scale MIs in this limit.
A promising method is based on differential equations (see
Ref.~\cite{Argeri:2007up} for a recent review) which provide the
expansion in an automatic way once the initial conditions are specified.
Let us as an example consider the five-line integral MI(5.2)
(cf. Fig.~\ref{fig::MI5}) which fulfills the following differential equation
\begin{eqnarray}
 \frac{{\rm d}}{{\rm d}z}\, \text{MI}(5.2) &=& 
\frac{d-4}{z-1}\, \text{MI} (5.2)
  +\frac{3 (z-2)(d-2)^2}{16m_t^6 (d-3) (z-4) (z-1)^2
   z}\,\text{MI}(2.2)\nonumber\\
&&\mbox{}
-\frac{(d-2)}{4 m_t^4
(z-1)^2}\,\text{MI}(3.4)-\frac{(d-2)}{8
   m_t^4 (z-1)^2}\,\text{MI}(3.5)\nonumber\\
&&\mbox{}
+\frac{(3 d-8) (z-6) }{16 m_t^4 (z-4) (z-1)
z}\,\text{MI}(3.10)\nonumber\\
&&\mbox{}
+\frac{(d-4)
   (z-9) (z+2)}{16m_t^2 (d-3) (z-4) (z-1)
   z}\,\text{MI}(3.10.1)\nonumber\\
&&\mbox{}
-\frac{\left(z^2-z+6\right)}{4 m_t^2 (z-4)
(z-1)^2 z}\,\text{MI}(3.10.2)+\frac{(d-3)
   }{2 m_t^2 (z-1)^2}\,\text{MI}(4.4)
\,.
\end{eqnarray}
MI(3.10.1) and MI(3.10.2) denote the MIs of the type 3.10 with a
squared massless and top quark propagator, respectively. In order to solve
this equation it is necessary to know the results of all integrals with
less than five lines. With the help of the ansatz
\begin{eqnarray}
  \text{MI}(5.2) = \sum c_{ijk}\,\epsilon^i z^j (\ln z)^k,
\end{eqnarray}
the differential equation can be expanded in $\epsilon$ and $z$. As a
result it reduces to algebraic equations for the
coefficients $c_{ijk}$. In every order in $\epsilon$ there is one constant
$c_{ijk}$ which can not be determined with this procedure. It it
obtained from the initial condition at $z=0$, which in the case of MI(5.2) can be
found in Ref.~\cite{Scharf:1993ds}. 
In this way we have computed expansion terms up to order $z^{10}$ which can be
found on the {\tt Mathematica} file mentioned above. For illustration we present the 
first two expansion terms of MI(5.2) which read
\begin{eqnarray}
  \text{MI}(5.2) m_t^2 &=&
\frac{1}{4} \pi ^2 \ln 2-2 \ln ^3 2-\frac{1}{3} \pi ^2 \ln 3+2 \ln ^2 2
\ln 3-\ln 2 \ln ^2 3+\frac{\ln ^3 3}{3}\nonumber\\
&&\mbox{}-\text{Li}_3(-2)+\frac{1}{2}
   \text{Li}_3\left(\frac{1}{4}\right)-2
   \text{Li}_3\left(\frac{2}{3}\right)+\text{Li}_3\left(\frac{3}{4}\right)+\frac{21 \zeta
(3)}{8}\nonumber\\
&&\mbox{}-i\pi\left(\frac{\pi^2}{12}+\frac{1}{2}\ln^2 2 \right)
- z\bigg[
i\pi\left(\frac{1}{2}-\frac{1}{4}\ln 2+\frac{3}{4}\ln 3-\frac{3}{8} \ln z
\right)\nonumber\\
&&\mbox{}+\frac{3}{8} \ln 3 \ln z-\frac{3 \ln ^2 3}{8}+\frac{1}{4} \ln 2 \ln
3-\frac{\ln^2 2}{4}-\ln 2\nonumber\\
&&\mbox{}+\frac{17 \pi
^2}{48}-\frac{1}{8}\text{Li}_2\left(\frac{3}{4}\right) \bigg] + {\cal O}(z^2)
\,.
\end{eqnarray}

Note that for some integrals the differential equation can be solved
with the help of Harmonic Polylogarithms~\cite{Remiddi:1999ew} which
immediately leads to a closed result.

It is interesting to mention that for the integrals MI(4.3) and MI(4.4) no
initial condition is needed in order 
to obtain all the coefficients in the ansatz. They are completely fixed by
the corresponding differential equation and the solutions for the integrals of the
subtopologies. For all other integrals initial conditions at $z=0$ are
required. As already mentioned above most of them can be found in the literature
or are quite simple to compute using standard techniques.
However, we could not get analytic results for five\footnote{One more
  coefficient can be 
  obtained analytically from the requirement that our final result is
  finite. It agrees perfectly with our numerical result.}
coefficients in the
$\epsilon$-expansion of the integrals MI(4.5), MI(4.8), MI(5.3) and MI(5.4)
at $z=0$. 
We calculated these coefficients using the Mellin-Barnes method (see, e.g.,
Ref~\cite{Smirnov:2004ym}) where we used the program packages {\tt AMBRE}~\cite{Gluza:2007rt}
and {\tt MB}~\cite{Czakon:2005rk}.

The Mellin-Barnes representation for a given integral is not
unique. In particular it might happen that the convergence of the 
resulting numerical integration turns out to be good in one case whereas 
a poor convergence is observed in other cases.
The crucial quantity in this respect is the asymptotic behaviour of the
$\Gamma$ function for large imaginary part which is given by
\begin{eqnarray}
  \Gamma(a\pm ib)\stackrel{\tiny b\to\infty}{\simeq}
  \sqrt{2\pi}e^{\pm i\frac{\pi}{4}(2a-1)}e^{\pm ib(\ln
    b-1)}e^{-\frac{b\pi}{2}}b^{a-\frac{1}{2}}
  \,,
  \label{eq::asymgamma}
\end{eqnarray}
where the first two exponential factors lead to oscillations.
Let us discuss this in more detail for the Mellin-Barnes representation
of the integral MI(4.5)
\begin{eqnarray}
 \text{MI}(4.5) &=& \left(\frac{e^{\gamma_E}
\mu^2}{m_t^2}\right)^{2\epsilon}
\int_{-i\infty}^{+i\infty} \frac{dz_1}{2\pi
i}\int_{-i\infty}^{+i\infty} \frac{dz_2}{2\pi i} \, 
 e^{i\pi(2 \epsilon+z_1)} 4^{-2 \epsilon-z_2}\nonumber\\
&&\hspace{-0em}\times  \frac{\Gamma
(1-\epsilon) \Gamma
   (-\epsilon-z_1+1) \Gamma (-z_1)\left[\Gamma (-2 
\epsilon-z_2+1)\right]^2 }{\Gamma (-3
   \epsilon-z_1+2)
   \Gamma (-2 \epsilon-z_1+2) \Gamma (-4 \epsilon-2
   z_2+2)}\nonumber\\
 &&\hspace{-0em}\times \Gamma
 (-4 \epsilon-z_1-z_2+2) \Gamma (z_1-z_2) \Gamma
 (\epsilon+z_2) \Gamma (2 \epsilon+z_2)
 \,,
 \label{eq::MB45}
\end{eqnarray}
where $\gamma_E=0.577216\dots$ and
the contour of integration is chosen in such a way that the poles of the
$\Gamma$ functions with $+z_i$ are separated from the poles of the
$\Gamma$ functions with $-z_i$. Using the package {\tt MB} we can expand the
integrand in $\epsilon$. For the finite 
contribution this leads to a sum of an analytic part, a one-dimensional
Mellin-Barnes integral and a two-dimensional one. The latter correspond to the
integral in Eq.~(\ref{eq::MB45}) for $\epsilon=0$. 
If we insert in this expression the asymptotic behaviour for the $\Gamma$
functions as given in Eq.~(\ref{eq::asymgamma}) one can
see that the integrand of the two-dimensional integral falls off
exponentially, except for $\mbox{Im}(z_2)=0, \mbox{Im}(z_1)<0$. 
On this line the drop-off only shows a power-law behaviour which is dictated by
the last factor of Eq.~(\ref{eq::asymgamma}). 
In our particular case the drop-off turns out to be extremely
slow for the integration contour chosen by {\tt MB} which corresponds
to $\mbox{Re}(z_1)=-1/4$ and $\mbox{Re}(z_2)=-1/2$. Thus it is hard
to get an accurate result by the numerical 
integration since a highly oscillating functions has to be integrated. 
A closer look to the fall-off behaviour in Eq.~(\ref{eq::asymgamma}) 
shows that it is possible to improve the drop-off for Im$(z_1)\to-\infty$
by taking residues of the integrand in $z_2$ and thus shifting
the integration contour for $z_2$ more and more to positive values for
Re$(z_2)$.
In this way the integrand becomes well-behaved and can be integrated
numerically with sufficiently high precision.

Let us mention that in the case of MI(4.5) there is an alternative possibility
to improve the numerical properties of the Eq.~(\ref{eq::MB45}): after the
variable transformation $z_2\to z_2-2\epsilon$ {\tt MB}  
chooses integration contours which lead to a rapid convergence of the numerical
integration. We have checked that both approached lead to the same results and
obtained 9 digits for the finite part of MI(4.5) in eleven minutes of CPU time.

The remaining three integrals show similar properties as MI(4.5).
In all cases it is possible to end up with integrals which could be
integrated numerically. Our results read
\begin{eqnarray}
 \text{MI}(4.5) &=& \left(\frac{\mu^2}{m_t^2}\right)^{2\epsilon} \bigg[
\frac{1}{2\epsilon^2} + \frac{1}{\epsilon}
\left(\frac{5}{2}-2\ln
   2+i\pi\right)
 -4.81543683(7)
 \nonumber\\&&\mbox{}
 + 4i\pi(1 - \ln2)
\bigg]\,, \nonumber\\
 \text{MI}(4.8)\vert_{z=0} &=& 
\left(\frac{\mu^2}{m_t^2}\right)^{2\epsilon}
\bigg[
\frac{1}{2\epsilon^2} + \frac{1}{\epsilon}
 \left(\frac{5}{2}+i\pi\right)+\frac{19}{2}-\frac{23 \pi 
^2}{24}-\frac{5 \ln^2 2}{2}
  -\frac{3 \ln 3}{2}\nonumber\\
   &&\mbox{}+\frac{5}{2} \ln 2 \ln 3-\frac{5}{4}
  \text{Li}_2\left(\frac{3}{4}\right) +\frac{i\pi}{2} (11 -5  \ln 2)
  - \epsilon \,16.690539(1)  \nonumber\\
 &&\mbox{} -i\pi\epsilon  \bigg(-\frac{45}{2}+\frac{35 \pi ^2}{24}+3 
\ln2-\frac{21
  \ln^2 2}{4}+6 \ln3  +2 \text{Li}_2(-2) \nonumber\\
 &&\mbox{}-4\text{Li}_2\left(\frac{1}{4}\right)\bigg)\bigg]\,,\nonumber\\
 m_t^2\,\text{MI}(5.3)\vert_{z=0} &=&  2.704628(4) - 5.167709(4)i\,,\nonumber\\
 m_t^2\,\text{MI}(5.4)\vert_{z=0} &=& 2.70543(6) - 1.91431(6)i
 \,.
\end{eqnarray}
The accuracy for the finite part of these integrals is sufficient to obtain
the final result with four
significant digits.

Note that contrary to the default settings of {\tt MB} we do not use
{\tt Vegas} for 
the multidimensional numerical integrations. Instead we use {\tt Divonne}
which is available from the Cuba library~\cite{Hahn:2004fe}. 
For the integrals we have considered it leads to
more accurate results using less CPU time. 

We have performed an independent check of the initial conditions for
all the MIs employing the method of sector decomposition.
In particular we used program {\tt FIESTA}~\cite{Smirnov:2008py}.


\section{\label{sec::gammatt}The $\gamma t\bar{t}$ vertex}

In this Section we discuss ${\cal O}(\alpha \alpha_s)$ corrections 
to the $\gamma t\bar{t}$ vertex due to  $W$ boson and gluon exchanges 
with incoming photon momentum at $q^2=4 m_t^2$, the production
threshold of top quark pairs. 
This leads to corrections to $h_{I,V}^{(1,1)}$ mediated by a virtual
photon, i.e., to the first term of $h_{I,V}^{\rm tree}$ in
Eq.(\ref{eq:hamp}).
We denote by $\Gamma_A^{\,t}$ the contribution of the sum of all
one-particle-irreducible diagrams to the $\gamma t\bar{t}$ vertex
and parameterize the radiative corrections in the form
\begin{eqnarray}
  \hat{\Gamma}_{A}^{\,t}=Q_t
  +{\hat{\Gamma}}_{A}^{\,t,\,(0,1)}
  +{\hat{\Gamma}}_{A}^{\,t,\,(1,0)}
  +\hat{\Gamma}_{A}^{\,t,\,(1,1)}
  \,,  
  \label{eq::Ghat}
\end{eqnarray}
where the hat denotes renormalized quantities.
Substituting $\hat{\Gamma}_{A}^{\,t}$ for
the $Q_t$ in first line of Eq.~(\ref{eq:hamp}) and retaining the relevant orders in the
electroweak and strong couplings leads to 
the corrections to the helicity amplitudes,
$h_{I,V}^{(0,1)}, h_{I,V}^{(1,0)}$ and $h_{I,V}^{(1,1)}$. 
We further decompose $\hat{\Gamma}_{A}^{\,t}$ (and similarly the quantities on the
right-hand side of Eq.~(\ref{eq::Ghat})) according to the contributions from
the Higgs, $Z$ and $W$ boson exchanges:
\begin{eqnarray}
  \hat{\Gamma}_{A}^{\,t}= \hat{\Gamma}_{A,H}^{\,t} + \hat{\Gamma}_{A,Z}^{\,t} +
  \hat{\Gamma}_{A,W}^{\,t} 
  \,.
\end{eqnarray}
In this paper we compute the corrections up to order $\alpha\alpha_s$ to
$\hat{\Gamma}_{A,W}^t$. The other two quantities have been
computed in Ref.~\cite{Eiras:2006xm} where the matching coefficients
$c_v^{\rm\, H,mix}$ and $c_v^{\rm\, Z,mix}$ have been introduced. We
have the following relations  
\begin{eqnarray}
  &&
  \hat{\Gamma}_{A,H}^{\,t,(1,1)} = \frac{\alpha\alpha_s}{\pi^2 s_w^2} \,
  Q_t C_F\, c_v^{\rm \, H,mix}, ~~~
  \hat{\Gamma}_{A,Z}^{\,t,(1,1)} = \frac{\alpha\, \alpha_s}{\pi^2 s_w^2}\,
  Q_t C_F\, c_v^{\rm \,Z,mix} 
  \,.
\label{eq:HandZmix}
\end{eqnarray}
In our calculation we adapt in the electroweak sector the
t'Hooft-Feynman gauge, i.e. $\xi_W=1$, which guarantees a simple form
of the $W$ boson propagator. Note that $\xi_W\not=1$ introduces an
additional mass scale in our calculation which would lead to
significantly more complicated integrals at two-loop order.
We want to mention that our final result depends on $\xi_W$. This
dependence only cancels out after including 
the self-energy and box contributions.
There is no gauge parameter dependence in 
$\hat{\Gamma}_{A,H}^{\,t}$ which only occurs in the vertex and
self-energy contributions.
Also the vertex corrections involving the $Z$ boson
($\hat{\Gamma}_{A,Z}^{\,t}$) are independent of the corresponding gauge
parameter.

Although the one-loop results are well-known, we start our discussion
from this order since they enter the 
renormalization of the two-loop expressions.


\subsection{One-loop corrections}

The renormalized QCD contribution is given by
\begin{eqnarray}
  \hat{\Gamma}_A^{\,t,\,(0,1)}
  &=&
  \Gamma_A^{\,t,\,(0,1)} + Q_t \, Z_2^{(0,1)}
  \,,
  \label{eq::GA1l}
\end{eqnarray}
where $Z_2$ is the onshell wave function renormalization for external
top quarks. 
The expressions on the right-hand side of Eq.~(\ref{eq::GA1l}) are given by
\begin{eqnarray}
  \Gamma_A^{\,t,\,(0,1)}
  &=&
  Q_t\,\frac{\alpha_s C_F}{4\pi}\,
  \bigg(\frac{\mu^2}{m_t^2}\bigg)^\epsilon\,
  \bigg[
    \frac{3}{\epsilon}
    -4
    +\epsilon\,\bigg(8+\frac{\pi^2}{4}\bigg)
    \bigg] ,
  \nonumber \\
  Z_2^{(0,1)}
  &=&
  \frac{\alpha_s C_F}{4\pi}\,
  \bigg(\frac{\mu^2}{m_t^2}\bigg)^\epsilon\,
  \bigg[
    -\frac{3}{\epsilon}-4
    -\epsilon\,\bigg(8+\frac{\pi^2}{4}\bigg)
    \bigg] ,
  \label{eq::GZ1l}
\end{eqnarray}
where $C_F=4/3$. In Eq.~(\ref{eq::GZ1l}) the ${\cal O}(\epsilon)$ 
terms are kept since they enter the finite part of the two-loop expression.
For the renormalized vertex we have the relation
\begin{eqnarray}
  \hat{\Gamma}_A^{\,t, (0,1)} 
  &=& 
  Q_t\,c_v^{(1)} 
  \, =\, 
  -8\, Q_t\,\frac{\alpha_s C_F}{4\pi}\left(1 + \epsilon \, L_\mu \right)
  +{\cal O}(\epsilon^2)
  \,,
\end{eqnarray}
where $L_\mu=\ln\frac{\mu^2}{m_t^2}$.
The one-loop formula for the electroweak corrections is given by 
\begin{eqnarray}
  {\hat{\Gamma}}_{A}^{\,t,\, (1,0)}
  &=&
  {\Gamma}_{A}^{t,\, (1,0)}
  +Q_t \, Z_{2}^{(1,0)}
  +T_t^{\,3}\, Z_{CT}^{(1,0)}
  \,,
\end{eqnarray}
where $Z_{CT}^{(1,0)}$ is a counterterm associated with $Z$-photon
mixing at zero-momentum transfer~\cite{Denner:1991kt} which reads
\begin{eqnarray}
  Z_{CT}^{(1,0)}
  &=&
  \frac{\alpha}{4\pi s_w^2}
  \left(-\frac{1}{\epsilon}-\ln\frac{\mu^2}{M_W^2}\right)+{\cal O}(\epsilon)
  \,.
\end{eqnarray}
As we will see later the ${\cal O}(\epsilon)$ term for $Z_{CT}^{(1,0)}$ 
is not needed.
We present the remaining two ingredients as a series 
expansion of $z=M_W^2/m_t^2$, the exact formulae are collected in
Appendix for convenience.
The results for $Z_{2,W}^{(1,0)}$ reads
\begin{eqnarray}
  Z_{2,W}^{(1,0)}
  &=&
  \frac{\alpha}{4\pi s_w^2}
  \bigg(\frac{\mu^2}{m_t^2}\bigg)^\epsilon\,
  \bigg\{
    -\frac{1}{\epsilon}\bigg(\frac{1}{4}+\frac{1}{8z}\bigg)
    - \frac{i \pi }{8 z}
    -\frac{L_{z}}{4}
    -\frac{i \pi}{2}
    -\frac{1}{4}
    -\left(\frac{L_{z}}{8}+\frac{i \pi }{8}+\frac{21}{16}\right) z
    \nonumber\\&&\mbox{}
    +\epsilon\,
    \bigg[
    \frac{7 \pi ^2}{96 z}
    +\frac{L_{z}^2}{8}
    -\frac{L_{z}}{4}
    +\frac{5 \pi ^2}{16}-\frac{i \pi}{2}-\frac{3}{4}
    +\bigg(
    \frac{L_{z}^2}{16}
    +\frac{3 L_{z}}{16}
    +\frac{\pi ^2}{12}
    -\frac{9 i \pi }{8}
    -\frac{27}{32}
    \bigg) z
    \bigg]
  \bigg\}
  \nonumber\\&&\mbox{}
  +{\cal O}(z^2)\,,
\end{eqnarray}
and ${\Gamma}_{A,W}^{\,t,\, (1,0)}$ is given by
\begin{eqnarray}
  {\Gamma}_{A,W}^{\,t,\, (1,0)}
  &=&
  \frac{\alpha}{4\pi s_w^2}\,Q_b\,
  \bigg(\frac{\mu^2}{m_t^2}\bigg)^\epsilon\,
  \bigg\{
    \frac{1}{\epsilon}\bigg(\frac{1}{8z}+\frac{1}{4}\bigg)
    +\frac{1}{z}\bigg(\frac{i\pi}{8}-\frac{2\ln{2}}{3}+\frac{1}{6}\bigg)
    \nonumber\\&&\mbox{}
    +\frac{L_z}{4}
    +\frac{i\,\pi}{2}
    -\frac{1}{12}
    -\frac{2 \ln{2}}{3}
    +\bigg(\frac{L_z}{8}+\frac{i\,\pi}{8}+\frac{2 \ln{2}}{3}+\frac{7}{48}\bigg)\,z
    \nonumber\\&&\mbox{}
  +\epsilon\, 
   \bigg[\,
    -\frac{1}{z}\bigg(\frac{7\pi^2}{96}+\frac{i\pi\big(4\ln{2}-1\big)}{6}
                       -\frac{2\ln{2}\big(3\ln{2}-5\big)}{9}-\frac{5}{18}
                \bigg)
    \nonumber\\&&\mbox{}
    -\frac{L_z^2}{8}
    +\frac{L_z}{4}
    -\frac{5\pi^2}{16}
    -\frac{i\pi\big(4\ln{2}-1\big)}{6}
    +\frac{2\ln{2}\,\big(3\ln{2}+1\big)}{9}
    +\frac{7}{36}
    \nonumber\\&&\mbox{}
    -\bigg(
    \frac{L_z^2}{16}
    +\frac{3 L_z}{16}
    +\frac{2\ln{2}\big(3\ln{2}+1\big)}{9}
    +\frac{\pi^2}{12} 
    -\frac{i \pi\big(16\ln{2}-1\big)}{24} 
    -\frac{115}{288}
    \bigg)\,z
   \bigg]
  \bigg\}
  \nonumber\\&&\hspace{-.7cm}\mbox{}
   +\frac{\alpha}{4\pi s_w^2}\,T_t^{\,3}\,
  \bigg(\frac{\mu^2}{m_t^2}\bigg)^\epsilon\,
  \bigg\{\,
    \frac{1}{\epsilon}\,
    \bigg(
    \frac{1}{4z}
    +\frac{3}{2}
    \bigg)
    +\frac{1}{z}\,
    \bigg(
    \frac{i\pi}{4}
    -\frac{2\ln{2}}{3}
    +\frac{2}{3}
    \bigg)
    \nonumber\\&&\mbox{}
    +\frac{3 i \pi}{2}
    -3 \ln{2}
    +3
    -\bigg(
    \frac{5 L_z}{8}
    -\frac{15}{16}
    +\frac{5i\pi}{8}
    -\frac{7\ln{2}}{4}
    \bigg)
    \, z
    \nonumber\\&&\mbox{}
    +\epsilon\,
    \bigg[\,
    -\frac{1}{z}\,
    \bigg(
    \frac{7\pi^2}{48}
    +\frac{2i\pi\big(\ln{2}-1\big)}{3}
    -\frac{2\ln{2}\big(3\ln{2}-8\big)}{9}
    -\frac{13}{9}
    \bigg)
    \nonumber\\&&\mbox{}
    -\frac{7\pi^2}{8}
    -3i\pi\big(\ln{2}-1\big)
    +3\ln{2}\big(\ln{2}-2\big)
    +6
    \nonumber\\&&\mbox{}
    +\bigg(
    \frac{5 L_z^2}{16}
    -\frac{11 L_z}{16}
    +\frac{5\pi^2}{12}
    +\frac{i\pi\big(7\ln{2}+1\big)}{4}
    -\frac{7 \ln^2{2}}{4}
    +\frac{39}{32}
    \bigg)
    \, z
               \bigg]
  \bigg\}
  \nonumber\\&&\mbox{}
  +{\cal O}(z^2) ,
\label{eq::Z2GA1l}
\end{eqnarray}
where $L_z=\ln{z}$ and $\alpha=e^2/(4\pi)$ is the fine-structure
constant in the Thomson 
limit. Again the ${\cal O}(\epsilon)$ terms are retained due to their relevance
for the two-loop renormalization. Note that we keep the imaginary parts of both 
$\Gamma_{A,W}^{\,t,\,(0,1)}$ and $Z_{2,W}^{(0,1)}$. At one-loop order this only
affects the finite part; at two loops also the pole parts are
concerned (see below).

\begin{figure}[t]
\begin{center}
\includegraphics[width=.4\textwidth]{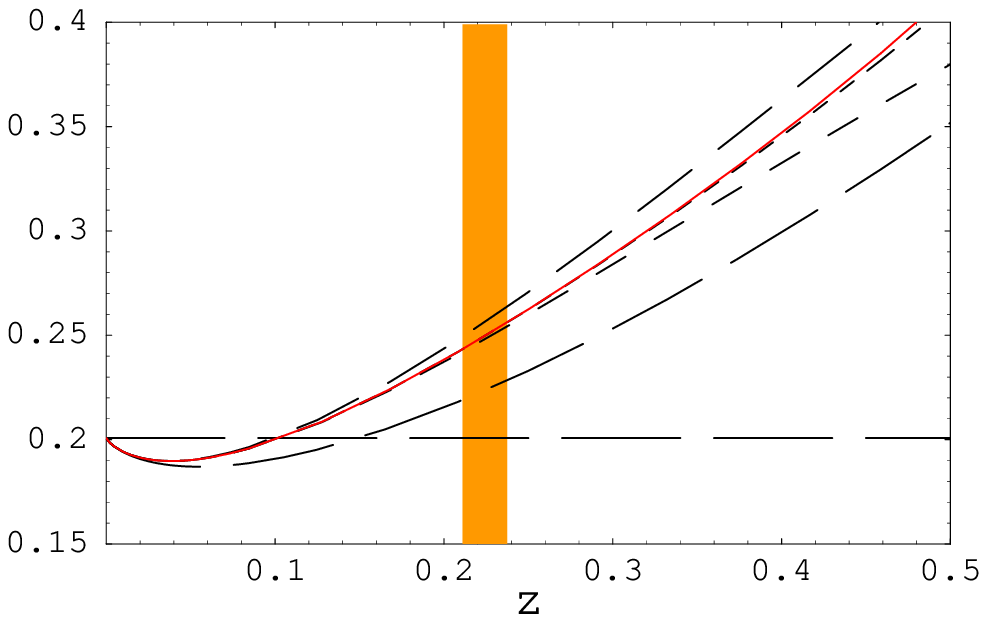}
\includegraphics[width=.4\textwidth]{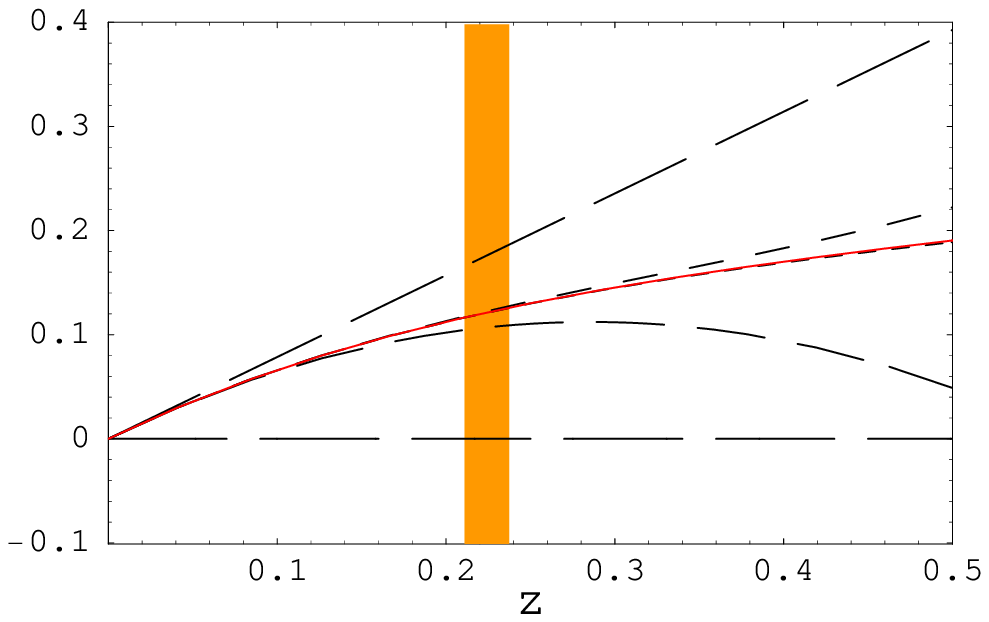}
\caption{Electroweak one-loop corrections to the photon vertex
  $\hat{\Gamma}_{A, W}^{\,t, (1,0)}$ normalized to $\alpha/(4\pi s_w^2)/z$ 
  as a function of
  $z=M_W^2/m_t^2$, for real part (left panel) and imaginary part (right
  panel), respectively.
  The (black) dashed lines include successively higher orders in 
  $z$ starting from $z^0$ (long dashes) to $z^4$ (short dashes) and
  the (red) solid line is the exact one-loop correction.
  The (orange) band marks the physical range of $z=(80.40~{\rm
  GeV}/m_t)^2$ varying the top quark mass between $165$ and $175$~{\rm GeV}.
  }
\label{fig::GA1l}
\end{center}
\end{figure}

We refrain from listing an analytical result for 
${\hat{\Gamma}}_{A}^{\,t,\, (1,0)}$ but compare in Fig.~\ref{fig::GA1l}
the approximated result to the exact one. The latter is
represented by the (red) solid line whereas the dashed lines correspond to
the expansions including successively higher
orders in $z$. As one can see, the expression including the
correction of order $z^3$ provides at the physical point $z\approx 0.23$ a
perfect approximation to the exact 
result far below the per cent level.
The approximated results are based on the following expressions
\begin{eqnarray}
  \hat\Gamma_{A,W}^{\,t,(1,0)}
  &=&
  \frac{\alpha}{4\pi s_w^2}
  \bigg[\,
    \frac{0.20}{z}
    +\big(0.48 +0.79\,i + 0.25 \ln{z}\big)
    \nonumber \\&&\mbox{}
    +z\,\big(-0.0024 - 1.37\,i - 0.44\,\ln{z}\big)
    +z^2\,\big(-0.072 + 1.39\,i + 0.44\,\ln{z}\big)
    \nonumber \\&&\mbox{}
    +z^3\,\big(0.34 - 0.53\,i - 0.17\,\ln{z}\big)
    +z^4\,\big(0.13 + 0.23\,i + 0.074\,\ln{z}\big)
    \bigg]
  \nonumber \\&&\mbox{}
  + {\cal O}(z^5)
  \nonumber \\
  &=&
  \bigg[2.48_{1/z}+0.25_{1}+0.39_{z}-0.095_{z^2}+0.017_{z^3}+0.00013_{z^4}
    \nonumber \\&&\mbox{}
    +i\,\big(2.11_{1}-0.80_{z}+0.18_{z^2}-0.015_{z^3}+0.0014_{z^4}\big) \,
    \bigg]\times 10^{-3}
  \,,
  \label{eq::num1l}
\end{eqnarray}
where the subscript in the last line indicate their order in the $z$
expansion and for the input parameters the following values have been
used~\cite{LEPEWWG,:2008vn}
\begin{eqnarray}
  \alpha_s=0.108\,,\qquad & \alpha(M_Z)=1/128.9\,, & \qquad 
  s_w^2=0.23\,,
  \nonumber\\
  M_W=80.40~\mbox{GeV}\,, &  m_t=172.4~\mbox{GeV}\,.
  \label{eq::num}
\end{eqnarray}
Note that in our numerical analysis we use $\alpha$ at high energy
  scale.\footnote{This is theoretically preferable because it is  
  devoid of non-perturbative hadronic effects.}

The corrections in Eq.~(\ref{eq::num1l})
are dominated by the leading terms 
proportional to $m_t^2/M_W^2$. 
One observes a rapid convergence, so that the term of order $z^3$
can safely be neglected. Inserting the results in Eq.~(\ref{eq::Ghat})
the overall size of the electroweak
corrections (from the diagrams involving a $W$ boson)
amounts to about 0.5\% and is unusually small.
For comparison, we note that $\hat\Gamma_{A,Z}^{\,t,(1,0)}$
and $\hat\Gamma_{A,H}^{\,t,(1,0)}$ lead to corrections of 0.3\% and 
3.2\% (for $M_H=120$~GeV), respectively.
Let us mention that the one-loop QCD corrections provides a
contribution ``$-61\times 10^{-3}$'' to the last line of
Eq.~(\ref{eq::num1l}) thus resulting in a 9\% correction.

From Eq.~(\ref{eq::num1l}) one obtains the corresponding 
corrections to the helicity amplitude as
\begin{eqnarray}
  \left[h_{I,V}^{(1,0)}\right]_{A, W} &=& Q_e \hat\Gamma_{A,W}^{\,t,(1,0)}
  \,,
  \label{eq::hiw1l}
\end{eqnarray}
which immediately leads to the correction to the cross section with
the help of Eq.~(\ref{eq:XSection}). Taking at tree-level both the photon and $Z$
exchange diagram we obtain a shift of 0.9\% to $R(e^+ e^-\rightarrow t\bar{t}X)$
due to the $W$ boson contribution to $\gamma t\bar{t}$ vertex at one-loop.


\subsection{Two-loop order $\alpha\alpha_s$ renormalization}

The renormalized $\gamma t\bar{t}$ vertex at order $\alpha\alpha_s$
is given by
\begin{eqnarray}
  \hat{\Gamma}_{A}^{\,t,\,(1,1)}
  &=&
  \Gamma_A^{\,t,\,(1,1)}
  +T_t^{\,3}\,  Z_{CT}^{(1,1)}
  +Q_t\, Z_2^{(1,1)}
  \nonumber \\
  &&
  +\left( Z_2^{(1,0)}\,\Gamma_A^{\,t,\,(0,1)}
  + Z_{2}^{(0,1)}\,\Gamma_A^{\,t,\,(1,0)}
  \right)
  + \frac{T_t^{\,3}}{Q_t}\, Z_{CT}^{(1,0)}\,\hat{\Gamma}_A^{\,t,\,(0,1)},
  \label{eq:the_renormalization_formula}
\end{eqnarray}
where the first line corresponds to genuine two-loop diagrams and the
second line consists of products of one-loop diagrams. 
For the latter we already listed all the relevant
expressions in the previous Subsection. 
Note that in the last term the renormalized one-loop
vertex appears and thus the ${\cal O}(\epsilon)$ term for
$Z_{CT}^{(1,0)}$ is not needed. 

Formula~(\ref{eq:the_renormalization_formula})
takes only care of the renormalization of the external lines and the
electric charge which means that the un-renormalized two-loop
quantities in the first line are understood as the sum of 
the amputated two-loop diagrams and the corresponding 
counterterm diagrams for the top quark mass and the top quark Yukawa
coupling. The latter are renormalized in the onshell scheme.

It is easy to see that for the two-loop counterterm we have
$Z_{CT}^{(1,1)}=0$ since at one-loop order only bosonic and no
fermionic diagrams contribute. 
The two-loop onshell wave function factor $Z_2^{(1,1)}$ has been
computed in Ref.~\cite{Eiras:2005yt}. We confirmed the result by
an independent calculation and added the imaginary part which is 
necessary in our framework. The result reads 
\begin{eqnarray}
  Z_{2,W}^{(1,1)}
  &=&
  \frac{\alpha}{4\pi s_w^2}
  \frac{\alpha_s C_F}{4\pi}\,
  \bigg\{\,
    \frac{1}{\epsilon^{\,2}}\bigg(\frac{3}{4}+\frac{3}{4 z}\bigg)
    +\frac{1}{\epsilon}
    \bigg[\,
    \frac{1}{z}\bigg(\frac{3 {L_\mu}}{2}+\frac{7}{4}+\frac{3i\pi }{8}\bigg) 
    +\frac{3{L_\mu}}{2}
    +\frac{3{L_z}}{4}
    \nonumber \\&&\mbox{}
    +\frac{17}{8}
    +\frac{3i\pi }{2}
    +\bigg(\frac{3{L_z}}{8}+\frac{63}{16}+\frac{3i \pi}{8}\bigg) z
    -\bigg(\frac{9\ln{z}}{4}-\frac{5}{4}+\frac{9i \pi}{4}\bigg) z^2
    \bigg]
    \nonumber \\&&\mbox{}
    +\frac{1}{z}
    \bigg(
    \frac{3 L_\mu^2}{2}
    +\frac{\big(14+3i\pi\big){L_\mu}}{4}
    +\frac{3 \zeta (3)}{2}
    -\frac{i\pi^3}{6}
    -\frac{\pi^2}{4}
    +\frac{27i \pi}{8}
    +\frac{79}{16}
    \bigg)
    \nonumber \\&&\mbox{}
    +\frac{3 L_\mu^2}{2}
    +\bigg(\frac{3{L_z}}{2}+\frac{17}{4}+3 i\pi \bigg) {L_\mu}
    -\frac{3 L_z^2}{8}
    +\frac{7 {L_z}}{4}
    -3 \zeta (3)
    +\frac{i\pi^3}{3}
    -\frac{\pi^2}{8}
    \nonumber\\&&\mbox{}
    +\frac{3i \pi }{2}
    -\frac{3}{16}
    +\bigg[\,\bigg(\frac{3{L_z}}{4}+\frac{3i\pi}{4}+\frac{63}{8}\bigg){L_\mu} 
    -\frac{3 L_z^2}{16}
    -\frac{\big(1+28i\pi\big){L_z}}{16}
    \nonumber\\&&\mbox{}
    -\frac{15 \zeta (3)}{2}
    +\frac{5i \pi^3}{6}
    +\frac{5 \pi^2}{6}
    -\frac{121i \pi }{24}
    +\frac{661}{32}
    \bigg]z
    +\bigg[\,
    \bigg(-\frac{9{L_z}}{2}+\frac{5}{2}-\frac{9i\pi}{2}\bigg){L_\mu}
    \nonumber\\&&\mbox{}
    +\frac{9 L_z^2}{8}
    +\bigg(\frac{1}{2}-\frac{5i\pi}{6}\bigg)\,L_z
    -\frac{317}{48}
    +\frac{547i\pi}{72}
    +\frac{79\pi^2}{36}
    -i \pi^3
    +9\zeta(3)
    \bigg] z^2
    \bigg\}
    \nonumber\\&&\mbox{}
    +{\cal O}(z^3)
  \,,
  \label{eq:2loop_Z2}
\end{eqnarray}
where terms up to order $z^2$ have been included ($\zeta(3)=1.20205\cdots$).

In the following we provide the result for the 
un-renormalized vertex corrections where the finite part is given in
numerical form. Our result reads
\begin{eqnarray}
  \Gamma_{A,W}^{\,t ,(1,1)}
  &=&
  \frac{\alpha}{4\pi s_w^2}
  \frac{\alpha_s C_F}{4\pi}\,
  \bigg\{
    \frac{2}{\epsilon^2}
    +\frac{1}{\epsilon}\,
    \bigg[
    -\frac{1}{z}\,\bigg(\frac{1}{3}-\frac{i\pi}{4}+\frac{\ln{2}}{3}\bigg)
    + 4 {L_\mu}
    -\frac{L_z}{4}
    +\frac{7i\pi}{4}
    +\frac{5}{3}
    -\frac{23\ln{2}}{6}
    \nonumber\\&&\mbox{}
    -z 
    \bigg(\,
    \frac{17{L_z}}{16}
    -\frac{47\ln{2}}{24}
    +\frac{17i\pi }{16}
    -\frac{121}{96}
    \bigg)
    - z^2 
    \bigg(\,
    \frac{17 {L_z}}{96}
    +\frac{185}{576}
    +\frac{17i\pi}{96}
    -\frac{65\ln{2}}{48}
    \bigg)
    \bigg]
    \nonumber\\&&\mbox{}
    +\frac{1}{z}
    \Big(
    \big[-2.66-3.79\big]
    -L_\mu\,\big[1.13-1.57i\big]
    \Big)
    +4.00\,L_\mu^2
    \nonumber\\&&\mbox{}
    +{L_\mu} 
    \big(-0.50 L_Z -\big[1.98-11.00i\big]\big)
    +0.13 L_z^2
    +0.08 L_z
    +\big[10.03-23.63i\big]
    \nonumber\\&&\mbox{}
    +z\,
    \bigg[
    L_\mu\,\Big(\big[5.24-6.68i\big]-2.13\,L_z\Big)
    +0.53\,L_z^2
    -L_z\,\big[0.77-7.76i\big]
    \nonumber\\&&\mbox{}
    -\big[11.84+5.62i\big]
    \bigg]
    +z^2\,
    \bigg[
    L_\mu\,\Big(\big[1.23-1.11i\big]-0.35\,L_z\Big)
    \nonumber\\&&\mbox{}
    +0.09\,L_z^2
    -L_z\,\big[3.80+1.16i\big]
    +\big[9.31+18.63i\big]
    \bigg]
    \bigg\}+{\cal O}(z^3)
  \,.
  \label{eq:the_bare_vertex}
\end{eqnarray}

Let us mention that in our calculation we allowed for a 
general QCD gauge parameter $\xi_S$ and used the independence of 
$\Gamma_A^{\,t ,(1,1)}$ as a welcome check for the correctness of our
result. Note that for the cancellation of $\xi_S$ it is important to
include the counterterm diagram for the top quark mass.
The remaining ingredients in
Eq.~(\ref{eq:the_renormalization_formula}) are individually
$\xi_S$-independent.
A further check of our calculation is based on a setup where we choose
$M_W=0$ from the very beginning. This leads to significantly simpler
expressions during the reduction to master integral, which is completely
independent from the one for finite $M_W$.


\subsection{${\cal O}(\alpha\alpha_s)$ corrections to the
  $\gamma t\bar{t}$ vertex} 

\begin{figure}[t]
\begin{center}
\includegraphics[width=.4\textwidth]{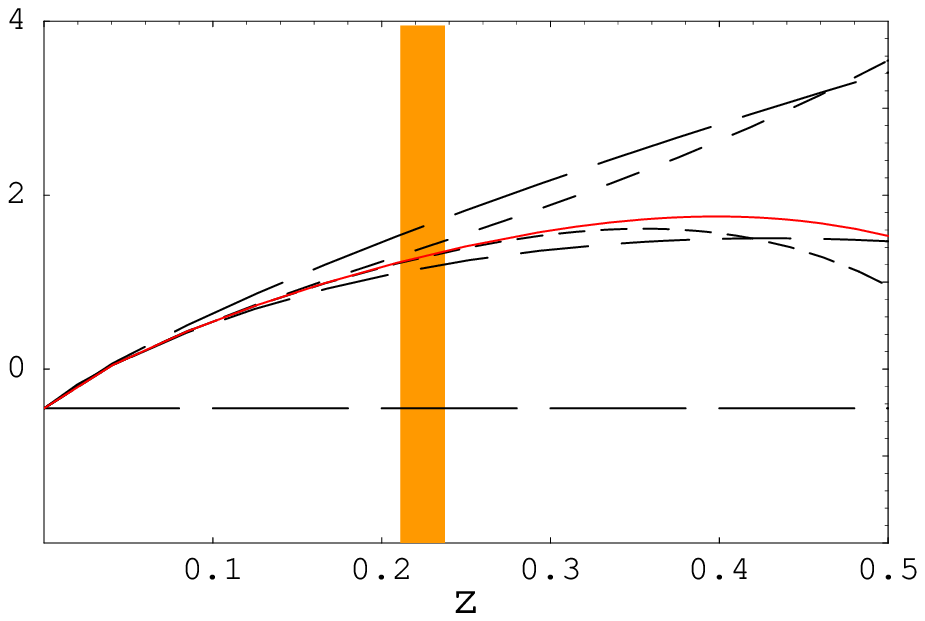}
\includegraphics[width=.4\textwidth]{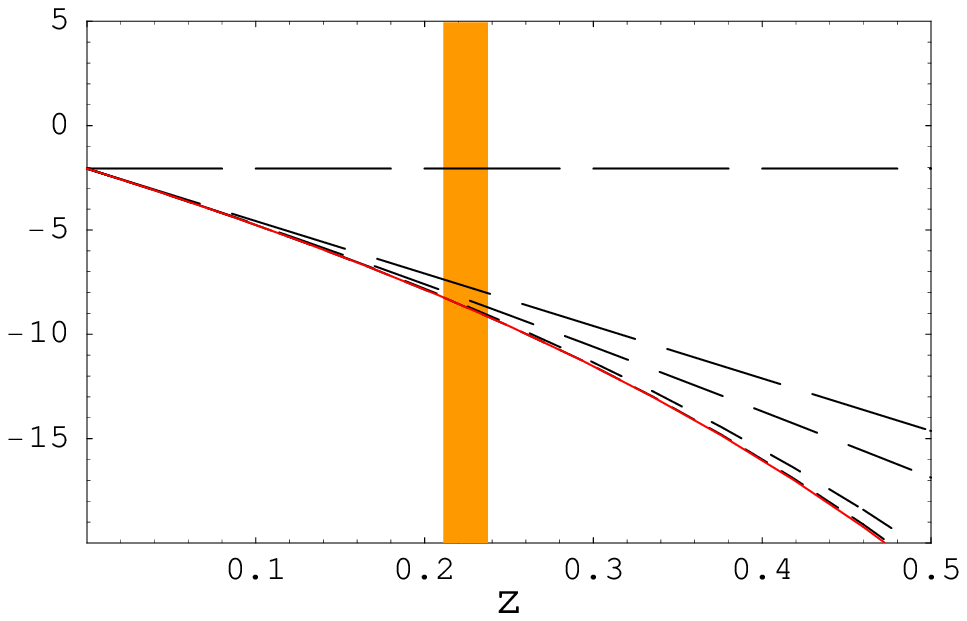}
\caption{Electroweak two-loop correction to the photon vertex
  $\hat{\Gamma}_{A, W}^{\,t, (1,0)}$ normalized to $(\alpha_s\alpha
  C_F)/(16\pi^2 s_w^2)/z$ as a function of
  $z=M_W^2/m_t^2$, for real part (left panel) and imaginary part (right
  panel), respectively.
  The (black) long-dashed lines include successively higher orders in 
  $z$ starting from $z^0$ (long dashes) to $z^4$ (short dashes) and
  the orange band is the physical range of $z$ (see Fig.~\ref{fig::GA1l}).
  }
\label{fig:GA2l}
\end{center}
\end{figure}

Inserting all ingredients into Eq.~(\ref{eq:the_renormalization_formula})
leads to 
\begin{eqnarray}
  \hat\Gamma_{A,W}^{\,t,(1,1)}
  &=&
  \frac{\alpha}{4\pi s_w^2}
  \frac{\alpha_s C_F}{4\pi}\,
  \bigg[\,
    \frac{1}{z}\,\bigg(-0.45-i\,2.06\bigg)
    +\bigg(6.34 -i\,25.14 - 2.00 \ln{z}\bigg)
    \nonumber \\&&\mbox{}
    +z\,\bigg(-6.27 - i\,6.10 + \big(2.16 + i\,4.10\big)\,\ln{z}\bigg)
    \nonumber \\&&\mbox{}
    +z^2\,\bigg(13.50 - i\,31.29 - \big(4.53 + i\,2.91\big)\,\ln{z}\bigg)
    \nonumber \\&&\mbox{}
    +z^3\,\bigg(-41.06 - i\,5.55 + \big(0.59 + i\,13.10\big)\,\ln{z}\bigg)
    \nonumber \\&&\mbox{}
    +z^4\,\bigg(17.11 - i\,9.99 - \big(1.86 + i\,5.36\big)\,\ln{z}\bigg)
    \bigg]
  +{\cal O}(z^5)
  \nonumber \\
  &=&
  \bigg[
    \big( -0.64_{1/z}+2.89_{1}-0.64_{z}+0.30_{z^2} -0.13_{z^3}+0.014_{z^4}\big)
    \nonumber \\&&\mbox{}
    +i\,\big(-2.91_{1/z}-7.73_{1}-0.83_{z}-0.39_{z^2}-0.081_{z^3}-0.0013_{z^4}\big) \,
    \bigg]\times 10^{-4}
    \,.
    \nonumber\\
  \label{eq::num2l}
\end{eqnarray}
In Fig.~\ref{fig:GA2l} we show
the result for $\hat{\Gamma}_A^{\,t,(1,1)}$
including first five terms of the $z$ expansion.
Taking the difference of two successive curves as a measure for the
quality of the approximation
we observe a rapid convergence at the physical point.
Note that in contrast to the one-loop case the leading $1/z$
contribution is numerically not dominant.

In analogy to Eq.~(\ref{eq::hiw1l}) we obtain for the correction to the
helicity amplitude
\begin{eqnarray}
  \left[ h_{I,V}^{(1,1)}\right]_{A, W} &=& Q_e \hat\Gamma_{A,W}^{\,t,(1,1)}
  \,,
  \label{eq::hiw2l}
\end{eqnarray}
which results in a correction of 0.1\% to $R$.\footnote{We did not
  take into account ${\cal O}(\alpha \alpha_s)$ interference terms, e.g.,
  $\delta R\sim (8\pi/s)\,\big[2 h_{I,V}^{(1,0)} h_{I,V}^{(0,1)} H_V\big]$.
  Such correction should be considered once the the two-loop box contributions
  are available.}

Let us in the following briefly compare the new vertex corrections to the 
ones induced by a Higgs and $Z$ boson. Note that the latter two contain a
non-trivial scale dependence which is canceled by the corresponding
contribution from the effective theory~\cite{Eiras:2006xm}. Choosing $\mu=m_t$
one obtains~\cite{Guth:1991ab,Eiras:2006xm}
\begin{eqnarray}
  &&
  \hat{\Gamma}_{A, H}^{t,(1,0)} = 21.1 \times 10^{-3} ~ (10.6\times 10^{-3} )
  ~ ~ ~\mbox{for}~~ M_H=120 ~ (200)~{\rm  GeV}\,,
  \nonumber \\ &&
  \hat{\Gamma}_{A, W}^{t,(1,0)}=3.0 \times 10^{-3}\,,
  \nonumber \\ &&
  \hat{\Gamma}_{A, Z}^{t,(1,0)}=1.7 \times 10^{-3}\,,
  \nonumber \\
  &&
  \hat{\Gamma}_{A, H}^{t,(1,1)}
  =-17.6 \times 10^{-3} ~ (-6.6\times 10^{-3} ) 
  ~ ~ ~\mbox{for}~~ M_H=120 ~ (200)~{\rm  GeV}\,,
  \nonumber \\ &&
  \hat{\Gamma}_{A, W}^{t,(1,1)}=0.2\times 10^{-3}\,,
  \nonumber \\ &&
  \hat{\Gamma}_{A, Z}^{t,(1,1)}=-1.0\times 10^{-3}\,.
  \label{eq::numG}
\end{eqnarray}
One observes quite small corrections from the $W$ and $Z$ boson
induced contributions. From Eq.~(\ref{eq::numG}) one can read off that
relatively big one-loop effects are
obtained for light Higgs boson masses. However, there is a strong cancellation
between the one- and two-loop terms resulting in corrections which have the
same size as the sum of the one- and two-loop contributions of the $W$ and $Z$
boson diagrams.
In general moderate effects are observed suggesting that in the
electroweak sector perturbation theory works well, which 
is in contrast to the pure QCD corrections.

Let us at this point comment on the imaginary parts contained in
Eqs.~(\ref{eq::num1l}) and~(\ref{eq::num2l}) which are not taken into
account in the numerical estimates for the corrections to $R$ presented
above. As we mentioned previously, ``${\rm Im}$'' in
Eq.~(\ref{eq:XSection}) applies to the imaginary 
part which corresponds to the $t\bar{t}X$ final state or 
experimentally indistinguishable cuts involving bottom quarks and $W$
bosons. Thus, it is necessary 
to separate imaginary parts arising from cutting, e.g., two $W$ boson
or two $b$ quark lines
from the $t\bar{t}X$ cuts in order to make a phenomenological prediction.
This requires a dedicated analysis in the loop calculation, which is 
beyond the scope of this paper (see, e.g., Ref.~\cite{Hoang:2004tg}).



\section{\label{sec::concl}Conclusions and outlook}

Mixed two-loop electroweak/QCD corrections to the $\gamma t\bar{t}$
vertex due to $W$ boson and gluon exchange have been computed.
The new contribution completes the order $\alpha\alpha_s$ corrections to the 
$\gamma t\bar{t}$ vertex.
The numerical evaluation leads to a shift of 0.1\% in the threshold production
cross section of top quark pairs at $e^+ e^-$ colliders which is small as
compared to the aimed 3\% uncertainty for the theory predictions.
Nevertheless, it is remarkable that in the sum of the order $\alpha$ and order
$\alpha\alpha_s$ correction terms the sizeable one-loop contribution from the
Higgs boson induced diagrams is screened resulting in numerical values
comparable to the $W$ and $Z$ boson contributions.

We want to mention that the corrections evaluated in this paper can be
taken over in a straightforward way to the vector coupling of the $Zt\bar{t}$
vertex. Note, that the axial-vector contribution is suppressed at threshold 
and thus one-loop corrections are sufficient.

The only missing building block in order to complete the order
$\alpha\alpha_s$ corrections to the process $e^+ e^-\to t\bar{t}$ 
are the two-loop box diagrams. They are technically more involved
and are thus postponed to future work.


{\bf Acknowledgements}

We would like to thank J.H. K\"uhn for helpful discussions and
A. Smirnov and  M. Tentyukov for providing the package {\tt FIESTA}
prior to its 
publication. Y.K. thanks N. Zerf for comparing the
electroweak one-loop corrections and Y. Sumino for discussions about
the implementation of the sector decomposition method within {\tt
  Mathematica}. D.S. acknowledges technical support by T. Hahn
concerning to the {\tt Cuba} library. This work is supported by the DFG
Sonderforschungsbereich/Transregio 9 ``Computergest\"utzte Theoretische
Teilchenphysik''.


\section*{Appendix: Exact result for $\Gamma_{A,W}^{\, t,(1,0)}$}

Keeping the full dependence on $\epsilon$
the exact one-loop result for the $\gamma t\bar{t}$ vertex due to the  
$W$ boson exchange reads (for massless bottom quarks)
\begin{eqnarray}
\Gamma_{A, W}^{\, t, (1,0)}
 &=&
\frac{\alpha}{4\pi s_w^2}
\bigg(\frac{1}{-3+2\,\epsilon}\bigg)
\bigg[
\big(Q_b+2 T_{t}^3\big)
\big(1+2(1-\epsilon) z\big)
\frac{A_0^{(\epsilon)}\big(M_W^2\big)}{8 M_W^2}
\nonumber \\ &&\mbox{}
+
\big(Q_b\,(5+z-4 \epsilon)+2 T_t^3\,(1+z)\big)
\frac{(1-z)\big(1+2(1-\epsilon) z\big)}{8z(1+z)}
B_0^{(\epsilon)}(m_t^2,M_W^2,0)
\nonumber \\ &&\mbox{}
-
Q_b \frac{(1+2(1-\epsilon) z)(1-\epsilon)}{z(1+z)}
B_0^{(\epsilon)}(4m_t^2,0,0)
\nonumber \\ &&\mbox{}
-
T_t^3 \,\bigg(\frac{1}{z}+5-4\epsilon\bigg)\, 
B_0^{(\epsilon)}(4m_t^2,M_W^2,M_W^2)
\bigg],
\end{eqnarray}
where $Q_b=-1/3$ is bottom quark electric charge normalized to the one
of positron. The corresponding contribution to the wave function
renormalization constant reads
\begin{eqnarray}
Z_2^{(1,0)}
&=&
\frac{\alpha}{4\pi s_w^2}
\bigg[
\big(1+2\,(1-\epsilon)\,z\big) 
\frac{A_0^{(\epsilon)}\big(M_W^2\big)}{8 M_W^2}
-
\frac{(1+z)\big(1+2\,(1-\epsilon)\, z\big)}{8z}
B_0^{(\epsilon)}(m_t^2,M_W^2,0)
\nonumber \\ &&\mbox{}
-
\frac{(1-z)\big(1+2\,(1-\epsilon)\,z\big)}{4z}\,
\big\{m_t^2\frac{\partial}{\partial m_t^2} B_0^{(\epsilon)}(m_t^2,M_W^2,0)\big\}
\bigg].
\end{eqnarray}
The loop-functions $A_0^{(\epsilon)}$ and $B_0^{(\epsilon)}$ are given by
\begin{eqnarray}
A_0^{(\epsilon)}(M_W^2)/M_W^2
&=&
-\bigg(\frac{e^{\gamma_E} \mu^2}{M_W^2}\bigg)^\epsilon \,
\Gamma(-1+\epsilon),
\label{eq::A0}
\\
B_0^{(\epsilon)}(4m_t^2,0,0)
&=&
\bigg(-\frac{e^{\gamma_E} \mu^2}{m_t^2}\bigg)^\epsilon \,
\frac{\sqrt{\pi}\,\Gamma(\epsilon)\,\Gamma(1-\epsilon)}{2\,\Gamma(\frac{3}{2}-\epsilon)},
\\
B_0^{(\epsilon)}(m_t^2,M_W^2,0)
&=&
\bigg(\frac{e^{\gamma_E} \mu^2}{m_t^2}\bigg)^\epsilon 
\bigg(-1+\frac{M_W^2}{m_t^2}\bigg)^{-\epsilon}\,
\frac{\Gamma(\epsilon)}{1-\epsilon}\,
\nonumber \\
&&
\times
{\,_2 F_1}\bigg(\epsilon,1-\epsilon,2-\epsilon;\,1/(1-M_W^2/m_t^2)\bigg),
\\
B_0^{(\epsilon)}(4m_t^2,M_W^2,M_W^2)
&=&
\frac{1}{2(1-\epsilon)}\,
\bigg(\frac{e^{\gamma_E} \mu^2}{2 M_W^2}\bigg)^\epsilon \,
\Gamma(\epsilon)\,
\nonumber \\
&&
\hspace*{-4cm}
\times
\bigg[
\bigg(1-\frac{1}{\sqrt{1-\frac{M_W^2}{m_t^2}}}\bigg)^\epsilon
\bigg(1+\sqrt{1-\frac{M_W^2}{m_t^2}}\bigg)
{_2  F_1\bigg(1-\epsilon,\epsilon,2-\epsilon;\frac{1}{2}\bigg(1+\frac{1}{\sqrt{1-\frac{M_W^2}{m_t^2}}}\bigg)\bigg)}
\nonumber \\
&&
\hspace{-4cm}
~~
+
\bigg(1+\frac{1}{\sqrt{1-\frac{M_W^2}{m_t^2}}}\bigg)^\epsilon
\bigg(1-\sqrt{1-\frac{M_W^2}{m_t^2}}\bigg)
{_2  F_1\bigg(1-\epsilon,\epsilon,2-\epsilon;\frac{1}{2}\bigg(1-\frac{1}{\sqrt{1-\frac{M_W^2}{m_t^2}}}\bigg)\bigg)}
\bigg],
\nonumber \\
\end{eqnarray}
where an analytic continuation by $(m_t^2+i 0)$ is understood. 
Expansions with respect to $\epsilon$ of the Gauss-hypergeometric
functions ${_2 F_1}$ around integer values is well known in the
literature (see, e.g., Ref.~\cite{Kalmykov:2006pu} or
the package {\tt HypExp}~\cite{Huber:2007dx}).




\begin{thebibliography}{99}

%
%

\bibitem{Martinez:2002st}
  M.~Martinez and R.~Miquel,
  Eur.\ Phys.\ J.\  C {\bf 27} (2003) 49
  [arXiv:hep-ph/0207315].

\bibitem{Hoang:2003xg}
  A.~H.~Hoang,
  Acta Phys.\ Polon.\  B {\bf 34} (2003) 4491
  [arXiv:hep-ph/0310301].

\bibitem{Pineda:2006ri}
  A.~Pineda and A.~Signer,
  Nucl.\ Phys.\  B {\bf 762} (2007) 67
  [arXiv:hep-ph/0607239].

\bibitem{Beneke:2008ec}
  M.~Beneke, Y.~Kiyo and K.~Schuller,
  arXiv:0801.3464 [hep-ph].

\bibitem{Beneke:2007pj}
  M.~Beneke, Y.~Kiyo and A.~A.~Penin,
  Phys.\ Lett.\  B {\bf 653} (2007) 53
  [arXiv:0706.2733 [hep-ph]];
  M.~Beneke and Y.~Kiyo,
  arXiv:0804.4004 [hep-ph] (Phys. Lett. B in press).

\bibitem{Guth:1991ab}
  R.~J.~Guth and J.~H.~K\"uhn,
  Nucl.\ Phys.\  B {\bf 368} (1992) 38.

\bibitem{Hoang:2006pd}
  A.~H.~Hoang and C.~J.~Reisser,
  Phys.\ Rev.\  D {\bf 74} (2006) 034002
  [arXiv:hep-ph/0604104].

\bibitem{Eiras:2006xm}
  D.~Eiras and M.~Steinhauser,
  Nucl.\ Phys.\  B {\bf 757} (2006) 197
  [arXiv:hep-ph/0605227].

\bibitem{Kniehl:1989yc}
  B.~A.~Kniehl,
  Nucl.\ Phys.\  B {\bf 347} (1990) 86.

\bibitem{Djouadi:1993ss}
  A.~Djouadi and P.~Gambino,
  Phys.\ Rev.\  D {\bf 49} (1994) 3499
  [Erratum-ibid.\  D {\bf 53} (1996) 4111]
  [arXiv:hep-ph/9309298].

\bibitem{Hoang:2004tg}
  A.~H.~Hoang and C.~J.~Reisser,
  Phys.\ Rev.\  D {\bf 71} (2005) 074022
  [arXiv:hep-ph/0412258].

\bibitem{Bodwin:1994jh}
  G.~T.~Bodwin, E.~Braaten and G.~P.~Lepage,
  Phys.\ Rev.\  D {\bf 51} (1995) 1125
  [Erratum-ibid.\  D {\bf 55} (1997) 5853]
  [arXiv:hep-ph/9407339].

\bibitem{Luke:1997ys}
  M.~E.~Luke and M.~J.~Savage,
  Phys.\ Rev.\  D {\bf 57} (1998) 413
  [arXiv:hep-ph/9707313].

\bibitem{Beneke:1997zp}
  M.~Beneke and V.~A.~Smirnov,
  Nucl.\ Phys.\  B {\bf 522} (1998) 321
  [arXiv:hep-ph/9711391].

\bibitem{Pineda:1997bj}
  A.~Pineda and J.~Soto,
  Nucl.\ Phys.\ Proc.\ Suppl.\  {\bf 64} (1998) 428
  [arXiv:hep-ph/9707481].

\bibitem{Brambilla:1999xf}
  N.~Brambilla, A.~Pineda, J.~Soto and A.~Vairo,
  Nucl.\ Phys.\  B {\bf 566} (2000) 275
  [arXiv:hep-ph/9907240].

\bibitem{Kniehl:2002br}
  B.~A.~Kniehl, A.~A.~Penin, V.~A.~Smirnov and M.~Steinhauser,
  Nucl.\ Phys.\  B {\bf 635} (2002) 357
  [arXiv:hep-ph/0203166].

\bibitem{Smirnov:2008pn}
  A.~V.~Smirnov, V.~A.~Smirnov and M.~Steinhauser,
  arXiv:0809.1927 [hep-ph] (Phys. Lett. B in press).

\bibitem{Fadin:1987wz}
  V.~S.~Fadin and V.~A.~Khoze,
  JETP Lett.\  {\bf 46} (1987) 525
  [Pisma Zh.\ Eksp.\ Teor.\ Fiz.\  {\bf 46} (1987) 417].

\bibitem{Penin:1998mx}
  A.~A.~Penin and A.~A.~Pivovarov,
  Phys.\ Atom.\ Nucl.\  {\bf 64} (2001) 275
  [Yad.\ Fiz.\  {\bf 64} (2001) 323]
  [arXiv:hep-ph/9904278].

\bibitem{Kniehl:1999ud}
  B.~A.~Kniehl and A.~A.~Penin,
  Nucl.\ Phys.\  B {\bf 563} (1999) 200
  [arXiv:hep-ph/9907489].

\bibitem{Manohar:2000kr}
  A.~V.~Manohar and I.~W.~Stewart,
  Phys.\ Rev.\  D {\bf 63} (2001) 054004
  [arXiv:hep-ph/0003107].

\bibitem{Kniehl:2002yv}
  B.~A.~Kniehl, A.~A.~Penin, M.~Steinhauser and V.~A.~Smirnov,
  Phys.\ Rev.\ Lett.\  {\bf 90} (2003) 212001
  [arXiv:hep-ph/0210161]; 
  Phys.\ Rev. \ Lett. \ {\bf 91} (2003) 139903, Erratum.

\bibitem{Hoang:2003ns}
  A.~H.~Hoang,
  Phys.\ Rev.\  D {\bf 69} (2004) 034009
  [arXiv:hep-ph/0307376].

\bibitem{Penin:2005eu}
  A.~A.~Penin, V.~A.~Smirnov and M.~Steinhauser,
  Nucl.\ Phys.\  B {\bf 716} (2005) 303
  [arXiv:hep-ph/0501042].

\bibitem{Czarnecki:1997vz}
  A.~Czarnecki and K.~Melnikov,
  Phys.\ Rev.\ Lett.\  {\bf 80} (1998) 2531
  [arXiv:hep-ph/9712222].

\bibitem{Beneke:1997jm}
  M.~Beneke, A.~Signer and V.~A.~Smirnov,
  Phys.\ Rev.\ Lett.\  {\bf 80} (1998) 2535
  [arXiv:hep-ph/9712302].

\bibitem{Marquard:2006qi}
  P.~Marquard, J.~H.~Piclum, D.~Seidel and M.~Steinhauser,
  Nucl.\ Phys.\  B {\bf 758} (2006) 144
  [arXiv:hep-ph/0607168].

\bibitem{Nogueira:1991ex}
  P.~Nogueira,
  J.\ Comput.\ Phys.\  {\bf 105} (1993) 279.

\bibitem{Harlander:1997zb}
  R.~Harlander, T.~Seidensticker and M.~Steinhauser,
  Phys.\ Lett.\ B {\bf 426} (1998) 125
  [hep-ph/9712228].

\bibitem{Seidensticker:1999bb}
  T.~Seidensticker,
  hep-ph/9905298.

\bibitem{PMDS}
  P.~Marquard and D.~Seidel,
  unpublished.

\bibitem{Laporta:1996mq}
  S.~Laporta and E.~Remiddi,
  Phys.\ Lett.\ B {\bf 379} (1996) 283
  [arXiv:hep-ph/9602417].

\bibitem{Laporta:2001dd}
  S.~Laporta,
  Int.\ J.\ Mod.\ Phys.\  A {\bf 15} (2000) 5087
  [arXiv:hep-ph/0102033].

\bibitem{Scharf:1993ds}
  R.~Scharf and J.~B.~Tausk,
  Nucl.\ Phys.\  B {\bf 412} (1994) 523.

\bibitem{Fleischer:1998dw}
 J.~Fleischer, F.~Jegerlehner, O.~V.~Tarasov and O.~L.~Veretin,
 Nucl.\ Phys.\  B {\bf 539} (1999) 671
 [Erratum-ibid.\  B {\bf 571} (2000) 511]
 [arXiv:hep-ph/9803493].

\bibitem{Seidel:2004jh}
  D.~Seidel,
  Phys.\ Rev.\  D {\bf 70} (2004) 094038
  [arXiv:hep-ph/0403185].

\bibitem{Kalmykov:2006pu}
 M.~Y.~Kalmykov,
 JHEP {\bf 0604} (2006) 056
 [arXiv:hep-th/0602028].

\bibitem{Piclum:2007an}
J.~H.~Piclum,
``Heavy quark threshold dynamics in higher order,''
Dissertation, Hamburg University, 2007.

\bibitem{Argeri:2007up}
  M.~Argeri and P.~Mastrolia,
  Int.\ J.\ Mod.\ Phys.\  A {\bf 22} (2007) 4375
  [arXiv:0707.4037 [hep-ph]].

\bibitem{Remiddi:1999ew}
  E.~Remiddi and J.~A.~M.~Vermaseren,
  Int.\ J.\ Mod.\ Phys.\  A {\bf 15} (2000) 725
  [arXiv:hep-ph/9905237].

\bibitem{Smirnov:2004ym}
  V.~A.~Smirnov,
  ``Evaluating Feynman Integrals,''
  Springer Tracts Mod.\ Phys.\  {\bf 211} (2004) 1.

\bibitem{Gluza:2007rt}
  J.~Gluza, K.~Kajda and T.~Riemann,
  Comput.\ Phys.\ Commun.\  {\bf 177} (2007) 879
  [arXiv:0704.2423 [hep-ph]].

\bibitem{Czakon:2005rk}
  M.~Czakon,
  Comput.\ Phys.\ Commun.\  {\bf 175} (2006) 559
  [arXiv:hep-ph/0511200].

\bibitem{Hahn:2004fe}
  T.~Hahn,
  Comput.\ Phys.\ Commun.\  {\bf 168} (2005) 78
  [arXiv:hep-ph/0404043].

\bibitem{Smirnov:2008py}
  A.~V.~Smirnov and M.~N.~Tentyukov,
  arXiv:0807.4129 [hep-ph].

\bibitem{Denner:1991kt}
  A.~Denner,
  Fortsch.\ Phys.\  {\bf 41} (1993) 307
  [arXiv:0709.1075 [hep-ph]].

\bibitem{LEPEWWG}
LEP Electroweak Working Group (LEP EWWG), 
see\\ {\tt http://lepewwg.web.cern.ch/LEPEWWG}

\bibitem{:2008vn}
   Tevatron Electroweak Working Group and CDF and D0
   Collaboration,
  arXiv:0808.1089 [hep-ex].

\bibitem{Eiras:2005yt}
  D.~Eiras and M.~Steinhauser,
  JHEP {\bf 0602} (2006) 010
  [arXiv:hep-ph/0512099].

\bibitem{Huber:2007dx}
  T.~Huber and D.~Maitre,
  Comput.\ Phys.\ Commun.\  {\bf 178} (2008) 755
  [arXiv:0708.2443 [hep-ph]].


\end{thebibliography}
\end{document}